# A Study on Modeling of Inputting Electrical Power of Ultra High Power Electric Furnace by using Fuzzy Rule and Regression Model

Choe Un-Chol, Yun Kum-Il, Kwak Son-Il

*Faculty of Electronics & Automation, Kim Il Sung University,*

*Ryongnam －Dong, Taesong District, Pyongyang, DPR Korea*

**Abstract:** In this paper a method to make inputting electrical model upon factors that affect melting process of high ultra power(UHP) electric furnace by using fuzzy rule and regression model is suggested and its effectiveness is verified with simulation experiment.

## 1.Introduction

One of the major means of iron production, UHP electric furnace is low of production cost and productivity of it is very high.

The problem of electrical power management in UHP electric furnace is an important problem associated with effective operation of it.

In UHP electric furnace, assuming that effective integrating power consume per ton of iron 250kwh~350kwh in melting process, inputting electrical power to UHP electric furnace is decided.

An important problem suggested in preceding study on electrode control of UHP electric furnace is control problem and modeling.

In reference[4] control method combining on-off control and PID control , in reference[5] improved PID control alghorithm, and, in reference[1], [2], [5], [9] control problem by using soft calculation technology including fuzzy logic are suggested.

Models that express dynamic characteristics is mentioned in references[4]~[7], complex phenomena are able to process by design methods about hybrid intellectual system accurately in references [10], [11].

In "non-serial fuzzy identification alghorithm" in reference [3], a modeling method of MISO non-linear dynamic system is described by using T-S fuzzy model.

In this paper a method to make input electrical model upon factors that affect melting process of UHP electric furnace by using fuzzy rule and regression model is suggested and effectivity is verified with simulation experiment.

## 2. Problem settings

In order to decide active power there are many considerable factors such as power factor, power transmission time, untransferable time , charge weight and mixing ratio of materials, and etc .

The model which has these factors as input, and active power as output, is typical nonlinear model.

This model can be expressed by fuzzy model with $m$ premise variables[2,3].

$$R^i : IF \ x_1(k) \ is \ A_1^i, \ x_2(k) \ is \ A_2^i, \cdots, x_m(k) \ is \ A_m^i$$
$$THEN \ y^i(k) = a_0^i + a_1^i x_1(k) + \cdots + a_m^i x_m(k) \quad (1)$$

In (1), $i = \overline{1,n}$

Given an input vector $(x_{1l}(k), x_{2l}(k), \cdots, x_{ml}(k))$, the output $\hat{y}_l(k)$ can be inferred by taking the weighted average of $y_l^i(k)$ for i=1, 2 , . . . , n.

The identification process of fuzzy model is the repeated operation of the calculation of optimal consequent and the correction of premise under the supposition that we know the structure and parameter of premise[2].

In this case variables of consequent of MISO linear equation are decided by variable elimination and parameters of equation are decided to minimize error between real measured output value and output value of fuzzy model in each step of variable elimination.

In premise we select the variables of consequent with premise variables and decide the number of variables by using fuzzy division of variable space.

Consequently, if fuzzy division by premise variable is poor, accuracy of approximation about non-linearity can be fall down terribly and the process of fuzzy modeling has many efforts.

On the other hand, approximation of complicated non-linearity is difficult by using the antecedent study result about regression linear equation [4].

In the identification of premise structure we select variables which equals consequent variable and decide number of variable by fuzzy division of variable space, but getting simulation result has many efforts.

So in the paper we select variables with large correlation coefficient about output as premise variables, and multidimensional nonlinear regression equation as consequent function and we set problem of identification of fuzzy rules with simplified method.

## 3. Premise variable decision using correlation coefficient

Accuracy of the rule model can be estimated by size of value of evaluation function.

$$J = \sqrt{\frac{1}{p} \sum_{k=1}^{p} (y_k - \hat{y}_k)^2} \quad (2)$$

Where $\hat{y}_k$ is output value of the fuzzy model at k th sampling instant, $y_k$ is output at k th sampling instant.

The procedure of premise structure identification by using variable elimination is as follows[2].
[algorithm 1]

1) A fuzzy model consisting of two fuzzy rules is constructed by first dividing the range of $x_1$ into {small} and {large} fuzzy subspaces:

$$R^1 : IF\ x_1\ is\ \{small\},\ THEN\ \cdots;$$
$$R^2 : IF\ x_1\ is\ \{large\},\ THEN\ \cdots;$$

Then, the premise parameters and the consequent structure and parameters are identified, and the $J$ of the model is calculated. Similarly, a fuzzy model dividing the range of each $x_i$ ($i = 1, 2, 3, \cdots, m$), is identified, and its $J$ is calculated.

Among the m models, the one with the smallest $J$ is determined. Its performance index and premise structure are denoted as $J(2)$ and STR(2), respectively.

2) Suppose that the premise variable of STR(2) is $x_i$ as shown in Fig.1(a).

At this step, each premise structure consists of three fuzzy subspaces : {small}, {medium}, and {large}.

There are two ways to construct a premise structure.

One is that the range of $x_i$ itself is divided into three fuzzy subspaces as shown in Fig. 1(b).

The other is that another variable $x_j$ is combined into STR(2) and its range is divided into two fuzzy subspaces as shown in Fig. 1(c) and (d).

For each premise structure which can be constructed in this step, a fuzzy model is identified and its $J$ is calculated. A fuzzy model with the smallest $J$ is found among these structures. Its $J$ and premise structure are marked as $J(3)$ and STR(3), respectively.

3) Suppose that we have obtained the premise structure STR(i—1) in step i—1 For i>3.

In step i, fuzzy models consisting of i fuzzy rules are constructed in two ways: either one of fuzzy subspaces of STR(i—1) is divided into two Fuzzy subspaces with respect to one of the premise variables of STR(i—1) as in Fig. 1(b) , or another new premise variable is combined into STR(i—1) and one of the fuzzy subspaces of STR(i—1) is divided with respect to the new premise variable just as in Fig. 1(c) and (d).

For each premise structure, a fuzzy model is identified and its $J$ is obtained. The one with the smallest $J$ is selected. Its $J$ and premise structure are denoted as $J(i)$ and STR(i), respectively.

4) If either of the following criteria is satisfied these arch will be stopped and the optimum premise structure for the model is obtained as STR(i). Otherwise, go to Step3.

-With a predetermined small value e>0, we have

$$\left| \frac{J(i) - J(i-1)}{J(i)} \right| < \varepsilon$$

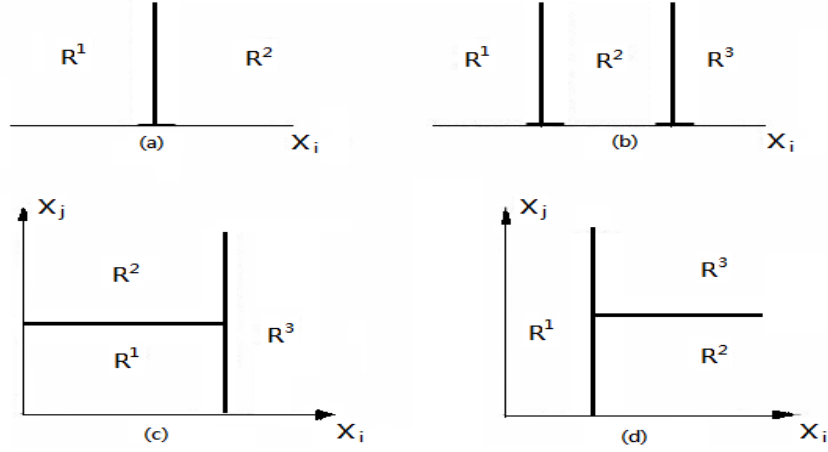

Fig. 1. A group of premise structure.

In variable elimination, all variables are dealt in from the beginning, but in this paper we use the method starting from the variable that has the largest correlation about output.

The proposed procedure is as follows.[algorithm 2]
1) We select variables that has the largest coefficient of association about output as premise variable.
2) We repeat ①~④ of algorithm 1.

## 4. Consequence function decision by using regression model

We selected regression model as consequence function as follows :

$$y(t) = a(q)^{-1}b(q)m(t) + e(t) = B(q)m(t) + e(t) \qquad (3)$$

Where $e(t)$ is white noise, $y(t)$ is output, $m(t)$ is input.

$$a(q) = 1 + a_1 q^{-1} + \cdots + a_{n_a} q^{-n_a}, \; b(q) = b_1 q^{-1} + \cdots + b_{n_b} q^{-n_b}$$

$$B(q) = [b_1(q)\, b_2(q) \cdots b_n(q)], \; m(t) = [m_1(t)\, m_2(t) \cdots m_n(t)]^T$$

In order to make variables of regression model as nonlinear, we select variables which has large coefficient of association about output, power or times of these variables as input of regression model. We decide parameters of regression model by using MATLAB.

## 5. Simulation and result analysis

**1) Simulation about regression model**

The factor which gives effect on the output $y$, active power needed per charge in UHP electric arc furnace, is expressed as $v_i, i = \overline{1,6}$ and calculated coefficient of association from samples is as follows.

| | Y | $v_1$ | $v_2$ | $v_3$ | $v_4$ | $v_5$ | $v_6$ |
|---|---|---|---|---|---|---|---|
| $v_1$ | -0.3889 | 1 | -0.7187 | -0.4664 | -0.2280 | 0.1134 | -0.1065 |
| $v_2$ | 0.5171 | | 1 | 0.0828 | 0.3250 | 0.0374 | 0.1187 |
| $v_3$ | 0.2089 | | | 1 | -0.1026 | -0.1785 | -0.0638 |
| $v_4$ | 0.4244 | | | | 1 | 0.0695 | 0.2505 |
| $v_5$ | 0.1607 | | | | | 1 | 0.1800 |
| $v_6$ | 0.3000 | | | | | | 1 |

Where $y$ is output, $v_1$ is quantity of charging material 1, $v_2$ is quantity of charging material 1, $v_3$ is quantity of charging material 1, $v_4$ is factor 1 which is affected by output, $v_5$ is factor 2 which is affected by output, $v_6$ is factor 3 which is affected by output.

In consideration of correlation coefficient, the variables of consequent are selected as follows.

$m_1 = v_1$, $m_2 = v_2$, $m_3 = v_1 \times v_4 + v_3$, $m_4 = v_1 \times v_2$, $m_5 = v_4$, $m_6 = v_5$, $m_7 = v_6$, $m_8 = v_1 \times v_5$, $m_9 = v_2 / v_1$

Simulation result and parameter identification result of regression model $y(t) = B(q)m(t)$ using all variables is as follows.

Where $B(q) = [b_1(q) \cdots b_9(q)]$, $m(t) = [m_1(t) \cdots m_9(t)]^T$.

$B(q) = [0.8541, -0.1796, -0.002624, 0.01023, 0.8787, 0.2436, 0.00966, -0.002499, 1.802]$

In a lot of simulation, the error between real measured value and calculation average value is 0.08~0.8, so modeling is poor.

**2) modeling simulation using fuzzy rule.**

Using variables which are large correlation coefficient about output, premise variables are made as follows.

$m_1 = v_1$, $m_2 = v_2$, $m_4 = v_1 \times v_2$, $m_5 = v_4$, $m_8 = v_1 \times v_5$, $m_9 = v_2 / v_1$

Descending order of correlation coefficient of these variables about output is $m_9, m_5, m_4$.

We selected these variables as premise variables and calculated performance function value.

As $m_9, m_5$ has largest correlation coefficient about output, these variables are selected as premise variables.

We normalize so that each factor has same importance and then eliminate variables one by one in consequent function.

Then completed modeling result is as follows.

R1: IF $m_9$ is S[0.08, 0.64] and $m_5$ is S[0.03, 0.88]    THEN   $y_1(t) = 4.954 + B_1(q)m(t)$

R2: IF $m_9$ is S[0.08, 0.64] and $m_5$ is B [0.04, 1.48]   THEN   $y_2(t) = 2.778 + B_2(q)m(t)$

R3: IF $m_9$ is B[0.18, 1.0] and $m_5$ is S[0.027, 0.88]    THEN   $y_3(t) = -240.499 + B_3(q)m(t)$

R4: IF $m_9$ is M[0.18, 1.0 1.7 2.16] and $m_5$ is B[-0.41, 1.48] THEN   $y_4(t) = 1.031 + B_4(q)m(t)$

R5: IF $m_9$ is B[1.8  3.0] and $m_5$ is B[0.14, 1.48]      THEN   $y_5(t) = 18.195 + B_5(q)m(t)$

Where  $m(t) = [m_1(t) \cdots m_9(t)]^T$,

$B_1(q) = [884.72, 19.55, 0, -7.36, 0, -42.947, 0.017, 0.37, 45.95]$

$B_2(q) = [-194.5, 0.173, 0, 1.4992, -16.67, 6.3754, 0.0113, 0.073, 5091.22]$

$B_3(q) = [384.4, 10.35, 0, 3.42, 0, -18.67, 0.029, 0.7536, 87.81]$

$B_4(q) = [2.5718, 0.0475, 0, 0.077, 4.3832, 0.3942, 0.0496, 0.0067, -88.6309]$

$B_5(q) = [-10.89, 0.39, 0, 0.284, 14.74, 3.076, 0.714, -0.1263, -15.8874]$

S=Small, M=Medium, B=Big.

Simulation result line is as follows.(Fig. 2)

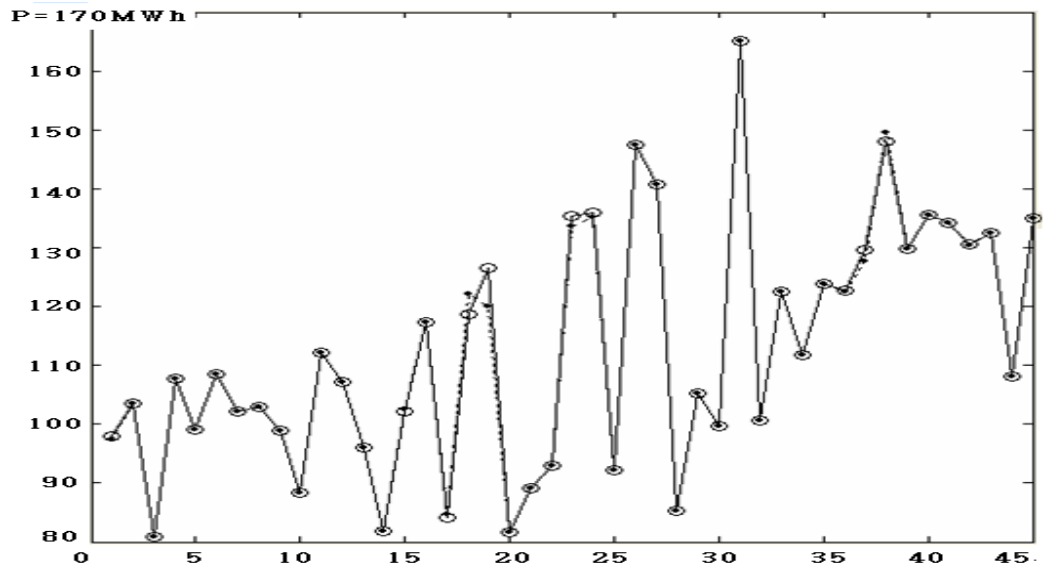

Fig.2. Simulation result using the proposed method

The relative error of simulation result using the proposed method has been extremely small as 0.33% and accuracy is higher than regression model as 99.7.

So model output was nearly same to real output.

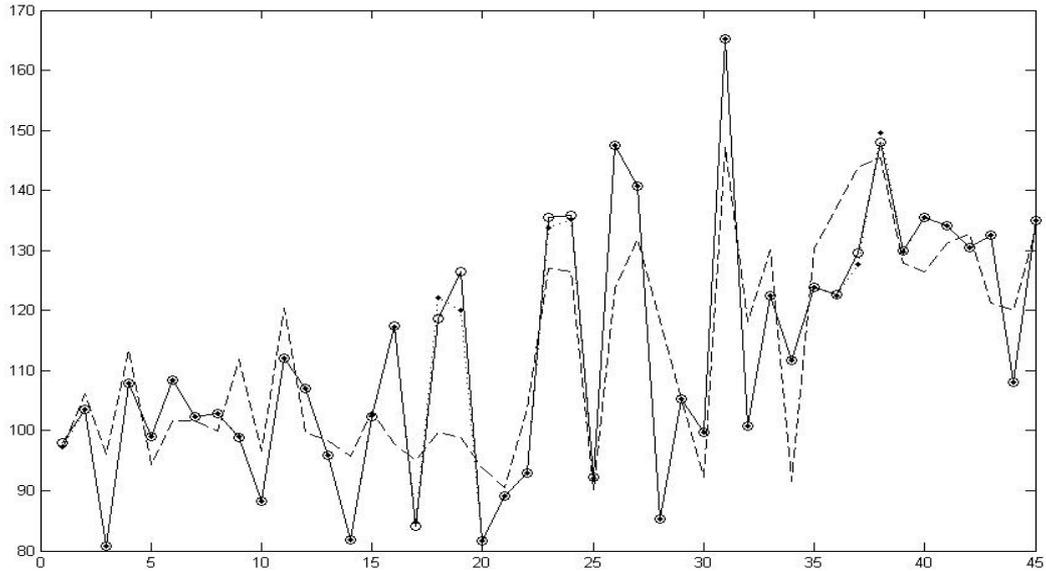

Fig.3. comparison curve between using regression model and using the proposed method.

Where real output is expressed as o and real line, output of regression model is expressed as point line, output of the proposed model is expressed as point line and black point.

## Conclusion

In this paper, we studied a method on the simplifying of fuzzy rule identification and validated effectiveness through the simulation with field operation data in UHP.

The proposed method can be widely used on modeling of various MISO nonlinear system.